
\font\twelverm=cmr10 scaled 1200    \font\twelvei=cmmi10 scaled 1200
\font\twelvesy=cmsy10 scaled 1200   \font\twelveex=cmex10 scaled 1200
\font\twelvebf=cmbx10 scaled 1200   \font\twelvesl=cmsl10 scaled 1200
\font\twelvett=cmtt10 scaled 1200   \font\twelveit=cmti10 scaled 1200
\font\twelvesc=cmcsc10 scaled 1200  
\skewchar\twelvei='177   \skewchar\twelvesy='60


\def\twelvepoint{\normalbaselineskip=12.4pt plus 0.1pt minus 0.1pt
  \abovedisplayskip 12.4pt plus 3pt minus 9pt
  \belowdisplayskip 12.4pt plus 3pt minus 9pt
  \abovedisplayshortskip 0pt plus 3pt
  \belowdisplayshortskip 7.2pt plus 3pt minus 4pt
  \smallskipamount=3.6pt plus1.2pt minus1.2pt
  \medskipamount=7.2pt plus2.4pt minus2.4pt
  \bigskipamount=14.4pt plus4.8pt minus4.8pt
  \def\rm{\fam0\twelverm}          \def\it{\fam\itfam\twelveit}%
  \def\sl{\fam\slfam\twelvesl}     \def\bf{\fam\bffam\twelvebf}%
  \def\mit{\fam 1}                 \def\cal{\fam 2}%
  \def\sc{\twelvesc}               \def\tt{\twelvett}
  \def\sf{\twelvesf}
  \textfont0=\twelverm   \scriptfont0=\tenrm   \scriptscriptfont0=\sevenrm
  \textfont1=\twelvei    \scriptfont1=\teni    \scriptscriptfont1=\seveni
  \textfont2=\twelvesy   \scriptfont2=\tensy   \scriptscriptfont2=\sevensy
  \textfont3=\twelveex   \scriptfont3=\twelveex  \scriptscriptfont3=\twelveex
  \textfont\itfam=\twelveit
  \textfont\slfam=\twelvesl
  \textfont\bffam=\twelvebf \scriptfont\bffam=\tenbf
  \scriptscriptfont\bffam=\sevenbf
  \normalbaselines\rm}



\def\beginlinemode{\endmode
  \begingroup\parskip=0pt \obeylines\def\\{\par}\def\endmode{\par\endgroup}}
\def\beginparmode{\endmode
  \begingroup \def\endmode{\par\endgroup}}
\let\endmode=\par
{\obeylines\gdef\
{}}
\def\singlespace{\baselineskip=\normalbaselineskip}

\def\oneandahalfspace{\baselineskip=\normalbaselineskip
  \multiply\baselineskip by 3 \divide\baselineskip by 2}
\def\doublespace{\baselineskip=\normalbaselineskip \multiply\baselineskip by 2}

\newcount\firstpageno
\firstpageno=2
\footline={\ifnum\pageno<\firstpageno{\hfil}\else{\hfil\twelverm\folio\hfil}\fi}
\def\toppageno{\global\footline={\hfil}\global\headline
  ={\ifnum\pageno<\firstpageno{\hfil}\else{\hfil\twelverm\folio\hfil}\fi}}
\let\rawfootnote=\footnote              
\def\footnote#1#2{{\rm\singlespace\parindent=0pt\parskip=0pt
  \rawfootnote{#1}{#2\hfill\vrule height 0pt depth 6pt width 0pt}}}
\def\raggedcenter{\leftskip=4em plus 12em \rightskip=\leftskip
  \parindent=0pt \parfillskip=0pt \spaceskip=.3333em \xspaceskip=.5em
  \pretolerance=9999 \tolerance=9999
  \hyphenpenalty=9999 \exhyphenpenalty=9999 }
\def\dateline{\rightline{\ifcase\month\or
  January\or February\or March\or April\or May\or June\or
  July\or August\or September\or October\or November\or December\fi
  \space\number\year}}
\def\received{\vskip 3pt plus 0.2fill
 \centerline{\sl (Received\space\ifcase\month\or
  January\or February\or March\or April\or May\or June\or
  July\or August\or September\or October\or November\or December\fi
  \qquad, \number\year)}}


\hsize=6.5truein
\vsize=8.9truein
\parskip=\medskipamount
\def\\{\cr}
\twelvepoint            
\doublespace            
\overfullrule=0pt       




\def\title                      
  {\null\vskip 3pt plus 0.2fill
   \beginlinemode \doublespace \raggedcenter \bf}

\def\author                     
  {\vskip 3pt plus 0.2fill \beginlinemode
   \singlespace \raggedcenter\sc}

\def\affil                      
  {\vskip 3pt plus 0.1fill \beginlinemode
   \oneandahalfspace \raggedcenter \sl}

\def\abstract                   
  {\vskip 3pt plus 0.3fill \beginparmode
   \singlespace ABSTRACT: }

\def\endtopmatter               
  {\endpage                     
   \body}

\def\body                       
  {\beginparmode}               

\def\head#1{                    
  \goodbreak\vskip 0.5truein    
  {\immediate\write16{#1}
   \raggedcenter \uppercase{#1}\par}
   \nobreak\vskip 0.25truein\nobreak}

\def\subhead#1{                 
  \vskip 0.25truein             
  {\raggedcenter {#1} \par}
   \nobreak\vskip 0.25truein\nobreak}

\def\beginitems{
\par\medskip\bgroup\def\i##1 {\item{##1}}\def\ii##1 {\itemitem{##1}}
\leftskip=36pt\parskip=0pt}
\def\enditems{\par\egroup}

\def\beneathrel#1\under#2{\mathrel{\mathop{#2}\limits_{#1}}}

\def\refto#1{$^{#1}$}           

\def\references                 
  {\head{References}            
   \beginparmode
   \frenchspacing \parindent=0pt \leftskip=1truecm
   \parskip=8pt plus 3pt \everypar{\hangindent=\parindent}}

\gdef\refis#1{\item{#1.\ }}                     

\gdef\journal#1, #2, #3, 1#4#5#6{               
    {\sl #1~}{\bf #2}, #3 (1#4#5#6)}            

\gdef\refa#1, #2, #3, #4, 1#5#6#7.{\noindent#1, #2 {\bf #3}, #4 (1#5#6#7).\rm}

\gdef\refb#1, #2, #3, #4, 1#5#6#7.{\noindent#1 (1#5#6#7), #2 {\bf #3}, #4.\rm}

\def\pr{\journal Phys.Rev., }

\def\rmp{\journal Rev.Mod.Phys., }

\def\cmp{\journal Comm.Math.Phys., }

\def\annp{\journal Ann.Phys.(N.Y.), }

\def\endreferences{\body}

\def\figurecaptions             
  {\endpage
   \beginparmode
   \head{Figure Captions}
}

\def\endpage                    
  {\vfill\eject}

\def\endpaper                   
  {\endmode\vfill\supereject}


\def\heading                            
  {\vskip 0.5truein plus 0.1truein      
   \beginparmode \def\\{\par} \parskip=0pt \singlespace \raggedcenter}

\def\subheading                         
  {\vskip 0.25truein plus 0.1truein     
   \beginlinemode \singlespace \parskip=0pt \def\\{\par}\raggedcenter}

\def\tag#1$${\eqno(#1)$$}

\def\align#1$${\eqalign{#1}$$}

\def\aligntag#1$${\gdef\tag##1\\{&(##1)\cr}\eqalignno{#1\\}$$
  \gdef\tag##1$${\eqno(##1)$$}}

\def\overset #1\to#2{{\mathop{#2}\limits^{#1}}}
\def\underset#1\to#2{{\let\next=#1\mathpalette\undersetpalette#2}}
\def\undersetpalette#1#2{\vtop{\baselineskip0pt
\ialign{$\mathsurround=0pt #1\hfil##\hfil$\crcr#2\crcr\next\crcr}}}


\def\ref#1{Ref.~#1}                     
\def\Ref#1{Ref.~#1}                     
\def\[#1]{[\cite{#1}]}
\def\cite#1{{#1}}
\def\(#1){(\call{#1})}
\def\call#1{{#1}}
\def\taghead#1{}
\def\frac#1#2{{#1 \over #2}}
\def\half{{\frac 12}}

\def\12{{1\over2}}

\def\sla{\raise.15ex\hbox{$/$}\kern-.57em}
\def\leaderfill{\leaders\hbox to 1em{\hss.\hss}\hfill}
\def\twiddle{\lower.9ex\rlap{$\kern-.1em\scriptstyle\sim$}}
\def\bigtwiddle{\lower1.ex\rlap{$\sim$}}
\def\gtwid{\mathrel{\raise.3ex\hbox{$>$\kern-.75em\lower1ex\hbox{$\sim$}}}}
\def\ltwid{\mathrel{\raise.3ex\hbox{$<$\kern-.75em\lower1ex\hbox{$\sim$}}}}
\def\square{\kern1pt\vbox{\hrule height 1.2pt\hbox{\vrule width 1.2pt\hskip 3pt
   \vbox{\vskip 6pt}\hskip 3pt\vrule width 0.6pt}\hrule height 0.6pt}\kern1pt}
\def\tdot#1{\mathord{\mathop{#1}\limits^{\kern2pt\ldots}}}

\def\pmb#1{\setbox0=\hbox{#1}%
  \kern-.025em\copy0\kern-\wd0
  \kern  .05em\copy0\kern-\wd0
  \kern-.025em\raise.0433em\box0 }

\catcode`@=11
\newcount\r@fcount \r@fcount=0
\newcount\r@fcurr
\immediate\newwrite\reffile
\newif\ifr@ffile\r@ffilefalse
\def\w@rnwrite#1{\ifr@ffile\immediate\write\reffile{#1}\fi\message{#1}}

\def\writer@f#1>>{}
\def\referencefile{
  \r@ffiletrue\immediate\openout\reffile=\jobname.ref%
  \def\writer@f##1>>{\ifr@ffile\immediate\write\reffile%
    {\noexpand\refis{##1} = \csname r@fnum##1\endcsname = %
     \expandafter\expandafter\expandafter\strip@t\expandafter%
     \meaning\csname r@ftext\csname r@fnum##1\endcsname\endcsname}\fi}%
  \def\strip@t##1>>{}}

\def\citeall#1{\xdef#1##1{#1{\noexpand\cite{##1}}}}
\def\cite#1{\each@rg\citer@nge{#1}}	

\def\each@rg#1#2{{\let\thecsname=#1\expandafter\first@rg#2,\end,}}
\def\first@rg#1,{\thecsname{#1}\apply@rg}	
\def\apply@rg#1,{\ifx\end#1\let\next=\relax
\else,\thecsname{#1}\let\next=\apply@rg\fi\next}

\def\citer@nge#1{\citedor@nge#1-\end-}	
\def\citer@ngeat#1\end-{#1}
\def\citedor@nge#1-#2-{\ifx\end#2\r@featspace#1 
  \else\citel@@p{#1}{#2}\citer@ngeat\fi}	
\def\citel@@p#1#2{\ifnum#1>#2{\errmessage{Reference range #1-#2\space is bad.}%
    \errhelp{If you cite a series of references by the notation M-N, then M and
    N must be integers, and N must be greater than or equal to M.}}\else%
 {\count0=#1\count1=#2\advance\count1 by1\relax\expandafter\r@fcite\the\count0,
  \loop\advance\count0 by1\relax
    \ifnum\count0<\count1,\expandafter\r@fcite\the\count0,%
  \repeat}\fi}

\def\r@featspace#1#2 {\r@fcite#1#2,}	
\def\r@fcite#1,{\ifuncit@d{#1}
    \newr@f{#1}%
    \expandafter\gdef\csname r@ftext\number\r@fcount\endcsname%
                     {\message{Reference #1 to be supplied.}%
                      \writer@f#1>>#1 to be supplied.\par}%
 \fi%
 \csname r@fnum#1\endcsname}
\def\ifuncit@d#1{\expandafter\ifx\csname r@fnum#1\endcsname\relax}%
\def\newr@f#1{\global\advance\r@fcount by1%
    \expandafter\xdef\csname r@fnum#1\endcsname{\number\r@fcount}}

\let\r@fis=\refis			
\def\refis#1#2#3\par{\ifuncit@d{#1}
   \newr@f{#1}%
   \w@rnwrite{Reference #1=\number\r@fcount\space is not cited up to now.}\fi%
  \expandafter\gdef\csname r@ftext\csname r@fnum#1\endcsname\endcsname%
  {\writer@f#1>>#2#3\par}}

\def\ignoreuncited{
   \def\refis##1##2##3\par{\ifuncit@d{##1}%
    \else\expandafter\gdef\csname r@ftext\csname r@fnum##1\endcsname\endcsname%
     {\writer@f##1>>##2##3\par}\fi}}

\def\r@ferr{\endreferences\errmessage{I was expecting to see
\noexpand\endreferences before now;  I have inserted it here.}}
\let\r@ferences=\references
\def\references{\r@ferences\def\endmode{\r@ferr\par\endgroup}}

\let\endr@ferences=\endreferences
\def\endreferences{\r@fcurr=0
  {\loop\ifnum\r@fcurr<\r@fcount
    \advance\r@fcurr by 1\relax\expandafter\r@fis\expandafter{\number\r@fcurr}%
    \csname r@ftext\number\r@fcurr\endcsname%
  \repeat}\gdef\r@ferr{}\endr@ferences}


\let\r@fend=\endpaper\gdef\endpaper{\ifr@ffile
\immediate\write16{Cross References written on []\jobname.REF.}\fi\r@fend}

\catcode`@=12

\citeall\refto		
\citeall\ref		%
\citeall\Ref		%

%
%
\def\D{\Delta}
\def\la{\langle}
\def\ra{\rangle}
\def\ria{\rightarrow}

\def\s{{\sigma}}
\def\a{\alpha}
\def\b{\beta}

\def\G{\Gamma}
\def\Tr{{\rm Tr}}
\def\ih{{ {i \over \hbar} }}
\def\trho{{\rho}}
\def\jjh{{j\_halliwell@vax1.physics.imperial.ac.uk}}
%
%
%
\centerline{\bf An Information-Theoretic Measure of Uncertainty}
\centerline{\bf due to Quantum and Thermal Fluctuations}
\vskip 0.3in
\author Arlen Anderson\footnote{$^{\dag}$}{E-mail address: arley@ic.ac.uk}
\vskip 0.2in
\centerline {and}
\vskip 0.1in
\author Jonathan J. Halliwell\footnote{$^{*}$}{E-mail address: \jjh}
\affil
Theory Group
Blackett Laboratory
Imperial College
South Kensington
London SW7 2BZ
UK
\vskip 0.5in
\centerline {\rm Preprint IC 92-93/25. April, 1993}
\vskip 0.2in
\centerline {\rm Submitted to {\sl Physical Review D}}.
\abstract
{
We study an information-theoretic measure of uncertainty for quantum
systems. It is the Shannon information $I$ of the phase space
probability distribution $\la z | \rho | z \ra $, where $|z \ra $
are coherent states, and $\rho$ is the density matrix. As shown by
Lieb and Wehrl, $I \ge 1 $, and this bound represents
a strengthened version of the uncertainty principle.
For a harmonic
oscillator in a thermal state, $I$ coincides with
von Neumann entropy, $- \Tr(\rho \ln \rho)$, in the high-temperature
regime, but unlike entropy, it is non-zero (and equal to the
Lieb-Wehrl bound) at zero temperature. It therefore supplies a
non-trivial measure of uncertainty due to both quantum and thermal
fluctuations. We study $I$ as a function of time for a class of
non-equilibrium quantum systems consisting of a distinguished system
coupled to a heat bath. We derive an evolution equation for $I$. For
the harmonic oscillator, in the Fokker-Planck regime, we show that
$I$ increases monotonically, if the width of the coherent states is
chosen to be the same as the width of the harmonic oscillator ground
state. For other choices of the width, and for more general
Hamiltonians, $I$ settles down to monotonic increase in the long
run, but may suffer an initial decrease for certain initial states
that undergo ``reassembly'' (the opposite of quantum spreading). Our
main result is to prove, for linear systems, that $I$ at each moment
of time has a lower bound $I_t^{min}$, over all possible initial
states. This bound is a generalization of the uncertainty principle
to include thermal fluctuations in non-equilibrium systems, and
represents the least amount of uncertainty the system must suffer
after evolution in the presence of an environment for time $t$.
$I_t^{min}$ is an envelope, equal, for each time $t$, to the time
evolution of $I$ for a certain initial state, which we calculate to
be a squeezed coherent state. $I_t^{min}$ coincides with the
Lieb-Wehrl bound in the absence of an environment, and is related
to von Neumann entropy in the long-time limit. The form of
$I_t^{min}$ indicates that the thermal fluctations become comparable
with the quantum fluctuations on a timescale equal to the
decoherence timescale, in agreement with earlier work of Hu and
Zhang. Our results are also related to those of Zurek, Habib and
Paz, who looked for the set of initial states generating the least
amount of von Neumann entropy after a fixed period of non-unitary
evolution.}
\endtopmatter

\head {\bf I. Introduction}

One of the most important features of quantum mechanics is the
uncertainty principle,
$$
 \D x \D p \ \ge \ { \hbar \over 2}
\eqno(1.1)
$$
Although frequently interpreted as a statement about the precision of
measurements, it may also be taken to mean that there is intrinsic
uncertainty in any phase space description of quantum systems. This
uncertainty may be especially significant for systems in certain states,
such as the ground state. However, in many quantum systems of interest
there is additional uncertainty due to thermal fluctuations, and moreover,
there may be regimes in which the thermal fluctuations dominate.
A number of questions then naturally arise: Is there a useful
measure of uncertainty due to both quantum and thermal
fluctuations? And, if so, what is the lower bound on this uncertainty,
analogous to (1.1)? What are the regimes in which each type of fluctuations
dominate? This paper addresses these questions.

Apart from being of interest in their own right, there are a number of
specific motivations for studying these issues. The principal one concerns
the general question of the emergence of classical behaviour in quantum
systems. Understanding this issue is one of the main aims of the decoherent
histories approach to quantum mechanics
[\cite{gellmann,griffiths,omnes,dowker}]. There (and in other
approaches [\cite{zurek,zurek2,zurekpazhabib,paz,zeh}]),
the process of decoherence is held to play an essential
role. This process typically occurs
as a result of interaction of the system under scrutiny with a wider
environment. But this same interaction also leads to essentially random
disturbances of the system, driving it off its classical path.
The probabilities for histories are typically found to be
peaked about classical histories, with some  width determined by quantum
effects and broadened by thermal fluctuations induced by interaction with
the environment [\cite{gellmann}]. It therefore becomes
important to gain a quantitative
understanding of both types of fluctuations, and to find the regimes in
which each are important.

In this paper we will explore an information-theoretic measure
of uncertainty due to both quantum and thermal effects, suitable
for the non-equilibrium quantum systems used in decoherence models.

We begin in Section II by describing the necessary background. We first
review some aspects of information theory. We then introduce a
quantum-mechanical phase space distribution. It is the distribution
$$
\mu(p,q) = \la z | \rho | z \ra
\eqno(1.2)
$$
where $\rho$ is the density matrix of the system, and $|z \ra $ are the
coherent states.  Our chosen measure of uncertainty is
the Shannon information $I$ of this distribution,
$$
I = - \int { dp dq \over 2 \pi \hbar} \ \mu(p,q) \ln \mu (p,q)
\eqno(1.3)
$$
As we shall explain, the uncertainty
principle manifests itself through the inequality,
$$
I \ \ge \ 1
\eqno(1.4)
$$
with equality if and only if $\rho$ is a coherent state [\cite{wehrl,lieb}].
Our main aim
is to generalize (1.4) to include the effects of thermal fluctuations
in non-equilibrium systems.

In Section III we study the properties of $I$ for a simple equilibrium
system -- the harmonic oscillator in a thermal state. This simple
example clearly illustrates how $I$ supplies a useful measure of both
thermal and quantum fluctuations. We then go on, in Section IV, to
consider non-equilibrium systems, the main topic of this paper. We
describe an important class of non-equilibrium systems consisting of a
distinguished system coupled to a heat bath (often referred to as open
quantum systems).

In Section V, we discuss the time-evolution of $I$ for non-equilibrium
systems. We show that $I_t$ generally settles down to monotonic
increase. There is, however, the possibility of an initial period of
decrease for specially chosen initial states which reassemble (the
opposite of wavepacket spreading).

In Section VI, we describe our main result. This is the demonstration
that $I_t$ has a non-trivial lower bound, the generalization of the
Lieb-Wehrl result (1.4) to include thermal fluctuations in non-equilibrium
systems. The function $I_t^{min}$ bounding $I_t$ from below is
generally not the time evolution of $I$ for some particular initial
state, but is an envelope. The initial state which achieves
$I_t^{min}$ at time $t$ (but generally not at any other time) is a
squeezed coherent state, with a specific value for the squeezing
factor depending on $t$.
$I_t^{min}$ is a
measure of the least amount of quantum and thermal noise the system
must suffer after non-unitary evolution for time $t$. The bound reduces
to the Lieb-Wehrl bound in the absence of an environment.

As we shall explain, there are three contributions to the uncertainty:

\item{(1)} There is the uncertainty intrinsic to quantum
mechanics, expressed through the uncertainty principle, (1.1).
This is not dependent on the dynamics. It is this uncertainty that
is referred to by the expression ``quantum fluctuations''.

\item{(2)} There is uncertainty that arises due to
the spreading or reassembly (the reverse of spreading)
of the wave packet. This effect depends on the dynamics,
and because quantum mechanics is time-symmetric, it may increase
or decrease the uncertainty.

\item{(3)} There is the uncertainty due to the coupling to a
thermal environment. This has two components: dissipation and diffusion
(the latter being responsible for the process of decoherence). This generally
tends to increase the uncertainty as time evolves.

\noindent The point is that the lower bound, $I_t^{min}$, includes
the effects (1) and (3), but avoids (2).

Finally, in Section VII, we summarize and discuss our results.  We
compare our results with calculations of Hu and Zhang [\cite{huzhang}], who
calculated the time evolution of the usual uncertainty function for
a particular initial state, and determined the timescale on which the
thermal fluctuations catch up with the quantum fluctuations.
We also compare with the results of Zurek,
Habib and Paz [\cite{zurekpazhabib,paz}], who looked for the set of initial
states which
generate the smallest amount of von Neumann entropy after a fixed
period of non-unitary evolution.

\head {\bf II. Background}

We now review the necessary background.

\subhead {\bf II(A). Information Theory}

Suppose one has a set of probabilities $p_i$ for a data set $S$ consisting of
discrete set of alternatives labeled by $i$, $i = 1,2 \cdots N $. One has $0
\le p_i \le 1$ and $ \sum_i p_i = 1 $. Then the Shannon information
of the data set is defined to be
$$
I(S) = - \sum_{i=1}^N p_i \ln p_i
\eqno(2.1)
$$
Here, $\ln$ is the logarithm to base $e$. $I(S)$ satisfies the inequalities
$$
0 \le I(S) \le \ln N
\eqno(2.2)
$$
It reaches its minimum if and only if $p_i = 1$, for one particular value of
$i$, and so $p_i = 0$ for all the other values. It reaches its maximum when
$p_i = {1 \over N} $ for all $i$. The information of a probability distribution
is therefore a measure of how strongly peaked it is about a given alternative.
For this reason, $I(S)$ is sometimes referred to as {\it uncertainty}, being
large for spread out distributions and small for concentrated ones.
This nomenclature is appropriate for purposes of this paper. The
expression (2.1) is also often referred to as the {\it entropy} of
the distribution, but we will not do so here, reserving the word
entropy for the von Neumann entropy of quantum statistical
mechanics (discussed in later sections).

In a similar manner for continuous distributions, let $X$ be a
random variable with probability density $p(x)$. Then
$\int dx \ p(x) = 1$. The information of $X$ is defined to be
$$
I(X) = - \int dx \ p(x) \ln p(x)
\eqno(2.3)
$$
Unlike the discrete case, $I(X)$ is no longer positive, since $p(x)$
is not a probability, but a probability density, so may be greater than $1$.
However, it retains its utility as a measure of uncertainty. This is
exemplified by a Gaussian distribution of variance $\D x$,
$$
p(x) = { 1 \over \left( 2 \pi (\D x)^2 \right)^{\half} } \
\exp \left( - { (x - x_0)^2 \over 2 (\D x)^2 } \right)
\eqno(2.4)
$$
It has information
$$
I(X) =
\ln \left( 2 \pi e (\D x)^2 \right)^{\half}
\eqno(2.5)
$$
{}From this we see that $I(X)$ is unbounded from below, and indeed,
approaches $- \infty$ as $\D x \rightarrow 0$ and $p(x)$ approaches a
delta-function. $I(X)$ is also unbounded from above, as may be
seen by taking the width $\D x$ to be very large. However, if the variance
is fixed, then a straightforward variational calculation shows that $I(X)$ is
maximized by the Gaussian distribution (2.4). We therefore have the
important inequality,
$$
I(X) \ \le \ \ln \left( 2 \pi e (\D x)^2 \right)^{\half}
\eqno(2.6)
$$

The generalization to probability distributions of more than one
variable is straightforward. For example, one has,
$$
I(X,Y) = - \int dx dy \ p(x,y) \ln p(x,y)
\eqno(2.7)
$$
and it is easy to show that
$$
I(X,Y) \ \le \ I(X) + I(Y)
\eqno(2.8)
$$
where $I(X)$ is the information of the distribution
$ \int dy p(x,y) $, and similarly for $I(Y)$. We also record another
useful result. Let $f(x), g(x) \ge 0$ and let $ \int dx g(x) = 1$.
Then
$$
- \left( \int dx f(x) g(x) \right)
\ln \left( \int dy f(y) g(y) \right) \ \ge
\ - \int dx f(x) g(x) \ln f(x)
\eqno(2.9)
$$
This is essentially due to the convexity of the function $x \ln x$,
and also holds in the discrete case.  Further details on information
theory may be found in the literature [\cite{info}].

\subhead {\bf II(B). Phase Space Distributions in Quantum Mechanics}

As stated above, our work is partly aimed at discussing the emergence of
classical behaviour. In this connection, it is often useful to introduce
quantum-mechanical phase space distributions. There are a variety of phase
space distributions that may be employed in quantum mechanics [\cite{wigner}].
In this paper we shall focus on the function,
$$
\mu (p,q) = \la z | \rho | z \ra
\eqno(2.10)
$$
where
$$
\la x |z \ra = \la x |p,q\ra = \left( {1\over 2 \pi \s_q^2}\right)^{1/4}
\exp \left(-{(x-q)^2\over 4\s_q^2} +i p x  \right)
$$
are the coherent states, with $ \s_p \s_q = \half \hbar $. We find it useful
to work with units with dimension, and for this reason it is necessary
to introduce the parameter $\s_q$ into the coherent state wave functions.
The function $\mu(p,q)$ is normalized
according to
$$
\int { dp dq \over 2 \pi \hbar} \ \mu(q,p)=1.
$$
It is readily shown that $\mu(p,q)$ is also equal to
$$
\mu (p,q) = 2 \int dp' dq' \ \exp
\left( - { (p-p')^2 \over 2 \s_p^2 } - { (q-q')^2
\over 2 \s_q^2 } \right) \ W_{\rho} (p',q')
\eqno(2.11)
$$
where
$ W_{\rho} (p,q) $ is the Wigner function of $\rho$, defined by
[\cite{wigner}],
$$
W_{\rho}(p,q) = { 1 \over 2 \pi \hbar }
\int d \xi \ e^{-{i \over \hbar} p \xi} \ \rho( q + \half \xi,
q - \half \xi )
\eqno(2.12)
$$
The distribution $\mu(p,q)$ is therefore a Wigner function, smeared
over an $\hbar$ sized region of phase space. This smearing renders
the distribution function positive, even though the Wigner function
is not in  general [\cite{jjh}]. The distribution (2.11) is
sometimes known as the Husimi distribution [\cite{husimi}], and has
appeared frequently in discussions of the Wigner function ({\it
e.g}, Refs.[\cite{gellmann,jjh,anderson}]).

The utility of the distribution function $\mu(p,q)$ will become apparent
as we expose some of its properties. We remark,
however, that $\mu$ is of the form,
$$
\mu (p,q) = \Tr \left[ P_z \rho \right]
\eqno(2.13)
$$
where $P_z = |z \ra \la z |$ is a coherent state projector (actually
only an approximate projector due to the over-completeness of the
coherent states). $\mu(p,q)$
therefore has the interpretation as the probability of a
simultaneous but approximate sampling of position and momentum.
Moreover, it may be shown that that by taking suitably weighted sums
over $p$ and $q$ of (2.13), an object of the form
$$
p({\bar x}_2, t_2, {\bar x}_1, t_1) =
\Tr \left[ P_{{\bar x}_2} (t_2) P_{{\bar x}_1}(t_1) \rho P_{{\bar x}_1}(t_1)
\right]
\eqno(2.14)
$$
may be obtained, where $ P_{\bar x}(t) $ denotes an imprecise position
sampling at time $t$. Eq.(2.14) is the probability for the history
characterized by the initial state $\rho$, and samplings of
position at times $t_1$ and $t_2$.
The distribution $ \mu(p,q)$ is therefore closely connected with the
the decoherent histories approach to quantum mechanics, which focuses
on objects of the form (2.14). In particular, it may be shown that
the degree to which expressions of the form (2.14) are peaked about
classical paths is limited by the degree of peaking of $\mu(p,q)$
in phase space. This is discussed in another paper
[\cite{jjh2}].

\subhead {\bf II(C). An Information-Theoretic Measure of Uncertainty}

We are interested in
the extent to which $\mu(p,q)$ is peaked about some
region of phase space. As we have discussed, the Shannon information is a
natural measure of the extent to which a probability distribution is peaked. We
shall therefore take as our measure of uncertainty, the information
$$
I(P,Q) = - \int { dp dq \over 2 \pi \hbar} \ \mu(p,q) \ln \mu(p,q)
\eqno(2.15)
$$

The uncertainty principle strongly suggests that a genuine phase
probability distribution in quantum mechanics cannot be arbitrarily
peaked about a  point in phase space. We therefore expect the
information (2.15) to possess a lower bound. Furthermore, since
coherent states are normally regarded as the states most
concentrated in phase space, we expect the lower bound to be the
value of $I$ on a coherent state. It turns out that both of these
expectations are true. It was conjectured by Wehrl [\cite{wehrl}], and proved
by Lieb [\cite{lieb}], that
$$
I(P,Q)  \ \ge \ 1
\eqno(2.16)
$$
with equality if and only if $ \rho $ is the density matrix of
a coherent state, $ |z' \ra \la z' | $.

The inequality (2.16) may be related to the usual uncertainty principle,
(1.1). One has the inequalities,
$$
\eqalignno{
\ln \left( { e \over \hbar} \D_\mu q \D_\mu p \right) & \ \ge \ I(Q) +I (P) \cr
& \ \ge \ I(P,Q)
&(2.17) \cr }
$$
The second inequality is an elementary property of information, (2.8);
the first is the inequality (2.6) applied to each of the marginal
distributions for $p$ and $q$,
where $ \D_{\mu} q $ and $\D_{\mu} p $ are the variances of the distribution
$ \mu(p,q)$ (the difference by a factor of $2 \pi \hbar $ is due to
our choice of phase space measure).
These variances are, however, not the quantum-mechanical
variances, since they include the variances of the coherent states.
Indeed, one has
$$
\eqalignno{
( \D_{\mu} q )^2 &= ( \D_{\rho} q )^2 + \s_q^2
&(2.18) \cr
( \D_{\mu} p )^2 &= ( \D_{\rho} p )^2 + \s_p^2
&(2.19) \cr }
$$
where $\D_{\rho}$ denotes the quantum-mechanical variance.
Now (2.16)-(2.19) together imply that
$$
\left( ( \D_{\rho} q )^2 + \s_q^2 \right)
\left(  ( \D_{\rho} p )^2 + \s_p^2 \right) \ \ge \hbar^2
\eqno(2.20)
$$
Now note that the width $\s_q $ in the coherent state is so far arbitary.
Minimizing (2.20) over $\s_q$ (and recalling that $\s_q \s_p = \half \hbar$),
we thus obtain the standard uncertainty relations, (1.1). An alternative
method of connecting the standard uncertainty relations with (2.16)
has been given by Grabowski [\cite{grabowski}].

Suppose now we have a state which is genuinely mixed. It may therefore be
written,
$$
\rho = \sum_n \ p_n \ |n \ra \la n |
\eqno(2.21)
$$
for some basis of states $|n \ra $, and where $p_n <  1 $.
One has
$$
\mu (p,q) = \sum_n \ p_n \ | \la  z | n \ra |^2
\eqno(2.22)
$$
The information of (2.22) will always satisfy (2.16), but this will be a
very low lower bound for a mixed state.  However, from the inequality
(2.9), one has
$$
\eqalignno{
I \ & \ge \ - \int {dp dq \over 2 \pi \hbar}
\ \sum_n \ | \la  z | n \ra |^2 \ p_n \ln p_n
\cr & = - \sum_n \ p_n \ln p_n
\cr & = - \Tr ( \rho \ln \rho ) \ \equiv S [\rho]
&(2.23) \cr}
$$
That is, $I$ is bounded from below by the von Neumann entropy, $ S[\rho]$.
As we shall see in the following section, this inequality can be close to
equality in the regime where thermal fluctuations are large.
This close connection with von Neumann entropy is one of the virtues of
our chosen measure of uncertainty, over other measures one might
contemplate ({e.g.}, the usual uncertainty function, $U = (\Delta_\rho q)^2
(\Delta_\rho p)^2 $).

{}From the above, we therefore see that $I$ is a useful of measure of
both quantum and thermal fluctuations. It possesses a lower bound
expressing the effect of quantum fluctuations, and is closely connected
to entropy, which in turn is a measure of thermal
fluctuations. In the following sections we will explore the further
properties of $I$, especially for non-equilibrium systems.

\head {\bf III. Fluctuations at Thermal Equilibrium }

To see some of the features of $I$ more clearly, consider the equilibrium case.
Let the density matrix be thermal, $\rho = Z^{-1} \ e^{- \beta H} $, where
$Z= \Tr (e^{-\beta H} )$ is the partition function, and $\beta = {1/ kT}$.
One has
$$
\la z | \rho |z \ra = {1 \over Z} \ \sum_n \ e^{- \beta E_n}
\ | \la z | n \ra |^2
\eqno(3.1)
$$
where $|n \ra$ are a set of energy eigenstates with eigenvalues $E_n$.
For simplicity, we restrict attention to the simple harmonic oscillator,
for which,
$$
H = \half \left( { p^2 \over M } + M \omega^2 q^2 \right)
\eqno(3.2)
$$
and so $E_n = \hbar \omega ( n + \half ) $, and
$$
| \la z | n \ra |^2 = { |z|^{2n} \over n! } \ e^{- |z|^2 }
\eqno(3.3)
$$
Here, $ z =\half \left( { q / \s_q}  + i { p / \s_p} \right)$,
where $\s_q \s_p = \half \hbar $, and we have made the choice
$ \s_q = ( \hbar /  2 M \omega)^{\half} $.
See Ref.[\cite{klauder}] for details about the coherent states.
One thus has
$$
\mu(q,p)=\la z | \rho  | z \ra   =
( 1 - e^{-\beta \hbar \omega})
\ \exp \left( - ( 1 - e^{-\beta \hbar \omega}) |z|^2 \right)
\eqno(3.4)
$$
The information (2.15) may then be computed explicitly. It is,
$$
I = 1 - \ln \left(1 - e^{-\beta \hbar \omega} \right)
\eqno(3.5)
$$

Eq.(3.5) is exactly the sort of result one would expect. As the
temperature goes to zero, $ \beta \ria \infty $, and the uncertainty
reduces to the Lieb-Wehrl result, (2.16), expressing purely quantum
fluctuations.  But for non-zero temperature, the uncertainty is larger,
tending to the value $- \ln \left( \beta \hbar \omega \right) $, as the
temperature goes to infinity. This limit expresses purely thermal
fluctuations.  For more general Hamiltonians, we expect the information
$I$ of the equilibrium thermal state to behave similarly (although we
have not been able to derive its explicit form).

It is of interest to compare (3.5) with the entropy,
$$
S = - \Tr ( \rho \ln \rho)
\eqno(3.6)
$$
The partition function is readily shown to be,
$$
Z = { 1 \over 2 \sinh ( \half \beta \hbar \omega ) }
\eqno(3.7)
$$
and the entropy is
$$
\eqalignno{
S &= - \beta { \partial \over \partial \beta } ( \ln Z ) + \ln Z
\cr & =
- \ln \left( 2 \sinh ( \half \beta \hbar \omega ) \right)
+ \half \beta \hbar \omega \coth ( \half \beta \hbar \omega )
&(3.8) \cr}
$$
For large temperatures (small $\beta$),
$$
S \approx \ - \ln \left( \beta \hbar \omega \right)
\eqno(3.9)
$$
$S$ therefore coincides with $I$ in the high-temperature limit,
On the other hand, $S \ria 0 $ as
the temperature goes to zero, whilst $I$ goes to a non-trivial lower bound.

We therefore see that $I$ is a useful measure of uncertainty, in both
the quantum and thermal regimes. Entropy, by contrast,
supplies a measure of uncertainty due only to thermal fluctuations. It
is therefore good in the thermal regime, but in the quantum regime, it
underestimates the intrinsic quantum uncertainty since it goes to zero
for pure states.

It is also useful to compare this measure of uncertainty with the more
standard measure,
$$
U = ( \Delta_{\rho} q)^2 \ (\Delta_{\rho} p)^2
\eqno(3.10)
$$
Here, $( \Delta_{\rho} q)^2$ is computed using $\la q^2 \ra = \Tr(q^2 \rho) $,
{\it etc.} One readily finds that
$$
( \Delta_{\rho} p)^2 \ = \ \omega^2 \ (\Delta_{\rho} q)^2
\eqno(3.11)
$$
Now Eq.(3.4) is product of Gaussians in $p$ and $q$, with variances
$ \Delta_\mu q $, $\Delta_\mu p$, say. The information of such
a distribution may be written,
$$
I = \ln \left( { e \over \hbar } \Delta_\mu q \Delta_\mu p \right)
\eqno(3.12)
$$
As in (2.18), (2.19), the variances of $q$ and $p$ in (3.12)
are not the same as the
quantum-mechanical variances, because they also include the
variances of the coherent state:
$$
\eqalignno{
(\Delta_\mu q)^2  & = (\Delta_{\rho} q)^2  + {\hbar \over 2 \omega}
&(3.13)\cr
(\Delta_\mu p)^2  & = (\Delta_{\rho} p)^2  + {\hbar \omega \over 2}
&(3.14) \cr}
$$
Inserting these in (3.12) and using (3.11), one obtains,
$$
I = \ln \left[ { e \over \hbar } \left(U^{\half} + \half \hbar \right) \right]
\eqno(3.15)
$$
This shows that, in this simple case,
there is a complete equivalence between $U$ and $I$ as measures of uncertainty.
We do not expect this equivalence to hold more generally, however.

Finally, we note that an information-theoretic uncertainty relation
including the effects of thermal fluctuations at thermal
equilibrium has been derived by Abe and Suzuki [\cite{abe}], using
thermofield dynamics. Their information-theoretic measure is
different to the one used here.

\head {\bf IV. Non-Equilibrium Systems}

Consider now the case of non-equilibrium systems, the main topic of
this paper. An important class of such systems in the present context
are those in which the total system naturally decomposes into a
distinguished system, ${\cal S}$ say, and the rest, summarily referred
to as the environment. ${\cal S}$ is then often referred to as an open
quantum system. One is interested only in the behaviour of ${\cal S}$,
and not in the detailed behaviour of the environment.  The
distinguished system is most completely described by the reduced
density matrix, $\rho$, obtained by tracing out over the environment.
The environment leaves its mark, however, in that the effective
evolution of the reduced density matrix alone is non-unitary.

A useful model of the type described above consists of a particle
moving in one-dimension in a  potential $V(x)$, linearly coupled to a
bath of harmonic oscillators in a thermal state. The environment is
characterized by a temperature $T$ and a dissipation coefficient $
\gamma $. This model has been the subject of many papers, so we will
give only the briefest of accounts here (for further details, see Refs.
[\cite{vernon,caldeira,dekker,unruh,hupaz,grabert}]).

After tracing out the environment, the reduced density matrix $\rho$
of the distinguished system evolves non-unitarily, according to the
relation
$$
\rho_t (x,y) = \int dx_0 dy_0 \ J(x,y,t|x_0,y_0,0) \ \rho_0 (x_0, y_0)
\eqno(4.1)
$$
Here, $J$ is the reduced density matrix propagator. It is given by
the path integral expression,
$$
J(x_f,y_f,t|x_0,y_0,0) = \int {\cal D}x {\cal D}y \ \exp \left(
\ih S[x] - \ih S[y] + \ih W[x,y] \right)
\eqno(4.2)
$$
where
$$
S[x] =  \int dt \left[ \half M \dot x^2 - V(x) \right]
\eqno(4.3)
$$
and $W[x(t),y(t)]$ is the Feynman-Vernon influence functional phase,
$$
\eqalignno{
W[x(t),y(t)] = & -
\int_0^t ds \int_0^s ds' [ x(s) - y(s) ] \ \eta (s-s') \ [ x(s') + y(s') ]
\cr &
+ i \int_0^t ds \int_0^s ds' [ x(s) - y(s) ] \ \nu(s-s') \ [ x(s') - y(s') ]
&(4.4) \cr }
$$
The explicit forms of the non-local kernels $\eta$ and $\nu$ may be found
in Refs.[\cite{hupaz,caldeira}]. We have assumed, as is typical in
these models, that the initial density matrix of the total system is
simply a product of the initial system and environment density matrices.

Considerable simplifications occur in a purely ohmic environment at high
temperature.  Take a regularized ohmic environment with cutoff frequency
$\Lambda$ having the spectral density
$$
C(\omega)= {2 M \gamma \omega\over \pi} e^{-\omega^2/\Lambda^2}.
\eqno(4.5)
$$
In the Fokker-Planck limit (see Ref.[\cite{hupaz}]),
one first takes the high temperature limit $\hbar/kT \ll \Lambda^{-1} $
and then lets the cutoff go to infinity, $\Lambda \rightarrow \infty$.
One finds
$$
\eqalignno{
\eta(s-s') &=  M \gamma \ \delta '(s-s')
&(4.6) \cr
\nu(s-s') &= { 2 M \gamma k T \over \hbar } \ \delta (s-s')
&(4.7) \cr }
$$
This limit is a simple and useful one, but our main results do not
depend on it.

The propagator $J$ may be evaluated exactly for the case of the
simple harmonic oscillator, $ V(x) = \half M \omega^2 x^2 $.
Introducing $ X = x+y $, $ \xi = x-y$, one has
$$
J(X_f, \xi_f, t  | X_0, \xi_0 , 0 ) = F^2 (t)
\ \exp\left(  \ih \tilde S - { \phi \over \hbar }  \right)
\eqno(4.8)
$$
where
$$
\tilde S= \tilde K(t) X_f \xi_f + \hat K(t) X_0 \xi_0
- L( t) X_0 \xi_f - N(t) X_f \xi_0
\eqno(4.9)
$$
and
$$
\phi  = A(t) \xi_f^2 + B(t) \xi_f \xi_0 + C(t) \xi_0^2
\eqno(4.10)
$$
Explicit expressions for the coefficients $\tilde K$, $\hat K$, $L$,
$N$, $A$, $B$ and $C$ are given
in Refs.[\cite{caldeira,hupaz}]. $F^2(t)=N/\pi$ is a
normalization factor, fixed
by imposing the condition
$$
\int dx dy \ \delta (x-y) \ J(x,y,t| x_0, y_0, 0 ) = \delta (x_0 - y_0)
\eqno(4.11)
$$
This ensures that ${\rm Tr} \rho_t = 1$ at all times. On the other
hand, tracing over the initial arguments of $J$ leads to
$$
\int dx_0 dy_0 \ \delta (x_0 - y_0) \ J(x,y,t|x_0,y_0,0) =
{ N \over L } \delta (x - y)
\eqno(4.12)
$$
We remark
that $\tilde S $ is in fact the action of the solution to the
boundary value problem for the harmonic oscillator with (non-local)
dissipation, for which the equation of motion is,
$$
\ddot X + \omega^2 X + 2 \int_0^s  ds' \ \eta(s-s') X(s') = 0
\eqno(4.13)
$$
In the classical limit, we expect that the quantum system reduces
to motion described by
this equation.

One may also derive an evolution equation for $\rho$, for general potentials.
Its most general form is [\cite{hupaz}],
$$
\eqalignno{
i \hbar { \partial  \trho \over \partial t} =
& - { \hbar^2 \over 2M } \left( { \partial^2 \trho \over \partial x^2 }
- { \partial^2 \trho \over \partial y^2 } \right)
+ \left[ V_R(x) - V_R(y) \right] \trho
\cr &
- i \hbar \Gamma (t) (x-y) \left( { \partial \trho \over \partial x}
- { \partial  \trho \over \partial y} \right)
- i \Gamma (t) h(t) (x-y)^2 \trho
\cr &
+ \hbar \Gamma(t) f(t) (x-y) \left( {\partial \rho \over \partial x}
+ {\partial \rho \over \partial y} \right)
&(4.14) \cr }
$$
Here $V_R(x)$ is the renormalized potential, $V_R(x) =V(x) + \half M
\delta \Omega^2(t) x^2 $. The explicit forms for the time-dependent
coefficients,
$ \delta \Omega(t)$, $\Gamma(t)$, $f(t)$, $h(t)$, are in general
rather complicated. Explicit expressions for them may be found in
Ref.[\cite{hupaz}]. In the Fokker-Planck limit, one has
$$
\Gamma(t) = \gamma , \quad h(t) = { 2 M kT \over \hbar}, \quad f(t) = 0
\eqno(4.15)
$$

The first two terms on the right-hand side of (4.14)
generate purely unitary
evolution (but with a renormalized potential).  The third term is
the dissipative term, and the fourth and fifth terms are diffusive
terms. In particular, the fourth term is responsible for the process
of decoherence discussed elsewhere
[\cite{zurek,zurek2,zurekpazhabib,paz,zeh}].

\def\p{\partial}

\head {\bf V. Time Evolution of $I_{{\lowercase{t}}}$}

We now study the evolution of $I$ as the density matrix $\rho$
evolves under the non-unitary evolution discussed in the previous
section. For simplicity, consider first the unitary evolution of $\rho$,
without an environment. One has
$$
\mu_t (p,q) = \la z | e^{-iHt} \rho_0 e^{iHt} | z \ra
\eqno(5.1)
$$
where $\rho_0$ is the density matrix at $t=0$, and may be pure or
mixed. The operators $ e^{\mp iHt} $, evolving $\rho_0$ forward in time,
may be equally regarded as evolving the coherent states backwards in time.
For a harmonic oscillator, the width $\s_q$ of the coherent states $|z \ra $
may be chosen to be the width of the ground state
(although this choice is by no means obligatory). With this choice,
the coherent states are preserved
under unitary evolution, with their centers following the
classical evolution:
$$
e^{-iHt} | p,q \ra = |  p_{cl}(t), q_{cl}(t) \ra
\eqno(5.2)
$$
The same is true for evolution backwards in time, with $t\rightarrow
-t$.  It is a standard result that the transformation from $(p,q)$ to
$(p_{cl}(t), q_{cl}(t))$ is a classical canonical transformation. The
effect of unitary evolution in (5.1) is therefore to perform a
canonical transformation on the arguments of $\mu_t(p,q)$ at $t=0$. It
is straightforward to see that our measure of uncertainty (2.15) is
invariant under canonical transformations of the variables of
integration. We therefore find that $I$ is constant under unitary
evolution for the harmonic oscillator, with the above special choice of
$\s_q$.

If the width $\s_q$ is not set to the above special value, then the
coherent states are not preserved under evolution by the harmonic
oscillator Hamiltonian. Likewise for more general Hamiltonians.  For
example, if the initial state is a coherent state, it will spread as
time evolves, and thus $I$ will increase from its initial value,
$I=1$.  Whether $I$ increases or decreases, however,
depends very much on the initial state. For example, the pure state $
e^{+iHt} |z\ra$, which could have a very large value of $I$, will
evolve under $e^{-iHt}$  into the coherent state $|z\ra$, possessing
the minimum value of $I$. This ``reassembly'' of a state sharply peaked
in phase space from a very spread out state will therefore cause $I$ to
decrease with time.

One would in fact expect initial states undergoing an initial decrease
of $I$ to be just as likely as ones undergoing an initial increase,
since quantum mechanics is a completely time-symmetric theory.
However, $I$ does in a certain sense capture the intuitive notion
that ``entropy increases'', even for pure states, in that it will
increase for initial states which might reasonably be described as
highly organized or special (namely, states that are sharply peaked
in phase space).

Now consider the coupling to an environment, as described in the
previous section. We shall derive an evolution equation for $I_t$.
We will first use the evolution equation for $\rho$, (4.13), to derive
an evolution for the Wigner function of $\rho$, (2.12).
Performing the Wigner transform of (4.13), one obtains,
$$
\eqalignno{
{ \p W \over \p t} = &- { p \over M} { \p W \over \p q} +
V_R^{\prime}(q)
{ \p W  \over  \p p } + 2 \Gamma(t) { \p \over \p p} (p W) + \hbar
\Gamma(t) h(t) { \p^2 W \over \p p^2}
+ \hbar \Gamma(t) f(t) { \partial^2 W \over \partial q \partial p}
\cr
& + \sum_{k=1}^{\infty} \left( {i \hbar \over 2 } \right)^{2k} { 1
\over (2k+1)!} V^{(2k+1)}(q) \ { \p^{2k+1} W \over \p p^{2k+1} }
&(5.3) \cr }
$$
The infinite power series incurred for general potentials makes
progress rather difficult. We shall therefore restrict attention to
the harmonic oscillator, $V(q) = \half M \omega^2 q^2 $, returning
at the end to a heuristic discussion  of the possible effects of
more general potentials. Now using the expression for
$\mu({\bar p},{\bar q})$, (2.11), one obtains,
$$
\eqalignno{
{ \p \mu \over \p t } =&- \ { \bar p \over M} { \p \mu \over \p \bar q}
+ M \omega^2_R(t) \bar q { \p \mu \over \p \bar p }
- \left( {\s_p^2 \over M} - M \omega^2_R(t)  \s_q^2 - \hbar
\Gamma(t) f(t) \right)
{ \p^2 \mu \over \p \bar p \p \bar q}
\cr &
+ 2 \Gamma(t) \mu
+ 2 \Gamma(t) \left( \bar p + \s_p^2 {\p \over \p \bar p} \right) {\p
\mu \over \p \bar p} + \hbar \Gamma(t) h(t) { \p^2 \mu \over \p \bar p^2}
&(5.4) \cr }
$$
Here, $\omega_R^2(t) = \omega^2 + \delta \Omega^2(t)$ is the
renormalized frequency.
Differentiating the expression for $I$, (2.15), one obtains, at some
length,
$$
\eqalignno{
\dot I = -2 \Gamma(t) & -
\left( {\s_p^2 \over M} - M \omega^2_R(t)  \s_q^2 - \hbar \Gamma(t) f(t)
\right)
\int {d{\bar p} d{\bar q} \over 2 \pi \hbar }
\ { 1 \over \mu} { \p \mu \over \p \bar p} { \p \mu \over \p \bar q}
\cr
& + \left( \hbar \Gamma(t) h(t)  + 2 \Gamma(t) \s_p^2 \right)
\int {d{\bar p} d{\bar q} \over 2 \pi \hbar}
\ {1 \over \mu} \left( { \p \mu \over \p \bar p} \right)^2
&(5.5) \cr}
$$
This is the exact result for the time evolution of $I$ for linear
systems.

Now the interesting question is whether we can say anything definite
about the monotonicity properties of $I$, given Eq.(5.5). First,
note that in the case of no environment, and for the harmonic
oscillator ({\it i.e.} $\omega \ne 0 $), it is possible to make the
choice
$$
\s_q^2 = { \hbar \over 2 M \omega}, \quad
\s_p^2 = \half M \omega \hbar
\eqno(5.6)
$$
and thus $\dot I = 0 $, as expected.

The next interesting case to consider is the Fokker-Planck limit,
(4.15), in which it is again useful to make the choice (5.6), and
the second term in (5.5) vanishes.
Consider the remaining terms in (5.5). The first term is $-2 \gamma$
and the coefficient of the last term is approximately $ 2M \gamma
kT$ (the $\s_p^2$ term is negligible in the Fokker-Planck limit).
Now the question is, what are the relative sizes of the first and last
terms in (5.5)?  Introduce the timescales,
$$
t_{dec} = { \hbar ^2 \over 2 M \gamma k T \s_q^2 }, \quad
t_{rel} = { 1 \over \gamma }
\eqno(5.7)
$$
The timescale $t_{dec}$ frequently emerges in studies of
decoherence, and is therefore called the decoherence timescale.  We
are not of course discussing decoherence {\it per se} here, but we
will use the nomenclature.  $t_{rel}$ is the relaxation timescale.
On  dimensional grounds it is clear that the first term will cause
$I$ to decrease on a timescale $t_{rel}$, and the last term will
cause it to increase on a timescale $t_{dec}$.  The important point
is that the relaxation time is typically very much longer than the
decoherence time [\cite{zurek2}], so the decoherence term will
dominate in (5.5). Thus
for the harmonic oscillator, with the choice (5.6), and in the
Fokker-Planck limit, $I$ will increase monotonically for any initial
state.

Now consider the case in which the choice (5.6) is not made. Closely
related is the case of the free particle, in which $\omega = 0$ in
(5.5), and $\s_p$ is arbitrary.  The question is whether $\dot I$ may
be rendered negative by the indefinite term in the integand (which is
associated with spreading or reassembly).  Physically, it is
reasonably clear how this may come about. As discussed above, it is
possible to choose special initial states that reassemble, at least
under unitary evolution, and will cause $I$ to decrease.  One would
expect to be able to identify a spreading or reassembly timescale,
$t_s$.  If the decoherence time scale is much shorter than the
spreading time scale, one would expect $I$ to increase monotonically,
since the environment acts before the system has time to undergo
reassembly. On the other hand, if the spreading time is shorter than
the decoherence time, an initial decrease may occur for carefully
chosen initial states, but this will eventually go over to increase
after a time of order $t_d$.
A similar situation could be expected to hold for more general
Hamiltonians.The Hamiltonian terms (in (5.3), say) may make $I$
increase or decrease, but eventually the diffusive terms will take
over and cause $I$ to increase.

These statements all apply to the high-temperature regime, in which
thermal effects will eventually dominate.  Eq.(5.5) is valid for all
regimes, and it would be of interest  to explore these, although we
do not do so here.

We now have a general picture of the behaviour of
$I$ under time evolution.  This sets the stage for the next Section,
in which we derive a lower bound on the behaviour of $I$.

Finally, we note that the analogue of Eq.(5.5) for von Neumann
entropy is very hard, if not impossible, to derive, even for linear
systems. Generally it can be obtained only if explicit
diagonalization of $\rho$ is possible, {\it e.g.}, for Gaussian
density matrices. For this reason, $I$ may be more practically
useful than $S$ as a measure of uncertainty, quite simply because it
is easier to calculate.

\def\p{{\bar p}}

\def\q{{\bar q}}

\head {\bf VI. A Lower Bound for $I_{{\lowercase{t}}}$}

We now come to the main point of this paper, which is to establish a
lower bound over all possible initial states for $I_t$, thus
generalizing (2.16) to include thermal fluctuations in time-evolving
non-equilibrium systems. We therefore seek a time-dependent function
$I_t^{min}$ such that for every time $t$,
$$
I_t \ \ge \ I_t^{min}
\eqno(6.1)
$$
$I_t^{min}$ represents the least amount of uncertainty the system must
suffer, after evolution for time $t$ in the presence of an
environment.  Clearly for consistency, we must have $I_t^{min} = 1 $
in the absence of an environment.

To fix ideas, consider first the case of no environment, for which the
evolution is unitary. The Lieb-Wehrl result is that the information
(2.15) at a fixed time is minimized by a system in a coherent state,
$\rho_0=|z'\ra\la z'|$. A harmonic oscillator initially in a coherent
state with a width given by Eq.(5.6)
evolves so that it remains in a coherent state, and therefore
$I_t= 1 = I_t^{min}$. It is easy to see that this
behavior is very special and cannot be realized for other Hamiltonians.
This is because Hamiltonian evolution generally does not preserve the
coherent states.  As described in the previous section, for every time
$\tau$, there is an initial state $e^{+iH\tau} |z' \ra $, with
non-minimal $I_t$ at $t=0$, which evolves to a coherent state at time
$\tau$, there minimizing $I_t$. After this, it disperses, and $I_t$ is
no longer minimal. $I_t$ is only minimized at $t=\tau$.

The implication of this is that $I_t^{min}$ is actually an envelope. No
particular $\rho_0$ realizes the minimum for all time -- instead there are a
succession of states which achieve the minimum. The minimum
$I_t^{min} = 1 $ is realized, at each time $t$, by the value of $I_t$
for the initial state $ e^{+iHt} |z' \ra $; that is, for the initial state
obtained by evolving the coherent state $|z'\ra$ at time $t$ backwards to
$t=0$.

Now consider the situation with an environment, as discussed in the previous
section. Instead of unitary evolution under $e^{-iHt}$, we now have
non-unitary evolution under the propagator $J$. As we have seen,
interaction with the environment will cause $I_t$ to increase in the long
run, but there is the possibility of an initial decrease of $I_t$, due to
the reassembly effect. We therefore expect $I_t^{min}$ to again be an
envelope: there will be many initial states which achieve $I_t^{min}$ for
some value of $t$, but there will be no initial state for which $I_t =
I_t^{min}$ for all $t$.

To find $I_t^{min}$, we will exploit the Lieb-Wehrl inequality (2.16).
It cannot, however,  be applied immediately to the case at hand. To see
why, consider again the case of no environment.  One is interested in
the information (2.15).  Application of the inequality (2.9) shows that
the minimum is achieved for a pure rather than mixed state.  One is
thus minimizing the integral
$$
I=-\int {dp dq \over 2 \pi \hbar }  \ |\la z| \psi\ra |^2
\ln |\la z| \psi\ra |^2
\eqno(6.2)
$$
over all square-integrable wavefunctions $\psi$.  The minimum is found to
be achieved for $| \psi \ra = |z' \ra$, a coherent state.  If one
expresses the state at a later time in terms of unitary evolution
from its initial value, $| \psi \ra=e^{-iHt}|\psi_0 \ra$, one
has the expression for the information at time $t$
$$
I_t=-\int {dp dq \over 2 \pi \hbar } \ |\la z| e^{-iHt}| \psi_0 \ra |^2
\ln |\la z | e^{-iHt}| \psi_0 \ra |^2 .
\eqno(6.3)
$$
Minimizing this over all square-integrable wavefunctions $| \psi_0 \ra$
is easy because $e^{-iHt}| \psi_0 \ra$ is itself a square-integrable
wavefunction, so the previous result applies, giving $| \psi_0 \ra =
e^{iHt} | z' \ra$, as discussed above.

Now we are interested in the more general case in which the propagator
is not unitary. We would like to know what the new lower bound on the
uncertainty is for systems that have undergone interaction with the
environment for time $t$.  Denoting the coherent state density matrix
by $\rho_z=| z \ra \la z |$, and the initial density matrix by
$\rho_0$, the information at time $t$ is given by
$$
I_t=-\int {dp dq \over 2 \pi \hbar }
\ \Tr\left( \rho_z J_t (\rho_0)\right)
\ln \Tr\left( \rho_z J_t (\rho_0)\right) .
\eqno(6.4)
$$
Here, $J_t(\rho_0)$ denotes the non-unitary evolution of $\rho_0$,
Eq.(4.1).  For each time $t$, we seek the $\rho_0$ that minimizes
(6.4).  Differently put, we need to minimize (6.4) over all density
matrices of the form $\rho_t= J_t (\rho_0)$, where $\rho_0$ is an
arbitrary density matrix. The feature that distinguishes this case from the
Lieb-Wehrl case discussed above is that this class of density matrices
is smaller than the class of {\it all} density matrices, since
evolution under $J$ is not invertible.  It is therefore difficult to
characterize the class over which to do the minimization.  Since
$J_t(\rho_0)$ is linear in $\rho_0$, and using the convexity property
(2.9), we again deduce that the minimizing $\rho_0$ must be pure. This
simplifies the problem somewhat, but the inconvenience stated still remains.

To get around this difficulty, we adopt the following strategy.
We are interested in the quantity,
$$
\eqalignno{
\mu_t (\p, \q) &= \la z | \rho_t | z \ra
\cr &= \int dx dy dx_0 dy_0 \ \la z | x \ra \la y | z \ra \ J
(x,y,t|x_0,y_0,0)
\ \rho_0 (x_0,y_0)
&(6.5) \cr}
$$
where $J$ is the reduced density matrix propagator. $\mu_t$ is then
conveniently written in the form,
$$
\eqalignno{
\mu_t(\p, \q) &= \int dx_0 dy_0 \ A_t^z (y_0,x_0) \ \rho_0 (x_0, y_0)
\cr & = {\rm Tr} \left( A_t^z \rho_0 \right)
&(6.6) \cr }
$$
where
$$
A_t^z (y_0,x_0) = \int dx dy \ \la z | x \ra \la y | z \ra \ J
(x,y,t|x_0,y_0,0)
\eqno(6.7)
$$
The quantity $A_t^z$ is therefore the final density operator
$|z\ra\la z|$ brought back from time $t$ to time zero using $J$.
Note, however, that $A_t^z$ is not a physical density matrix, since
from (4.12), $\Tr A_t^z = {N \over L}$ (although one does have
$ \int { d \p d \q \over 2 \pi \hbar} A_t^z = 1 $).
Using the Wigner representation, (2.12), one may write
$$
\mu_t (\p, \q) = 2 \pi \hbar \int dp dq \ W_{A_t^z} (p,q) \ W_{\rho_0}(p,q)
\eqno(6.8)
$$
Since $J$ is Gaussian for the linear case considered here,
$A_t^z (x_0,y_0)$ and $W_{A_t^z}(p,q)$ are also Gaussian.

Compare this to the Lieb-Wehrl result, (2.16). The latter may be
regarded as stating that the information of the distribution
$$
\mu (p, q) =
2 \int dp' dq'
\ \exp \left( - { (p' - p)^2 \over 2\sigma_p^2 }
-{ (q' - q)^2 \over 2\sigma_q^2 } \right)
\ W_{\tilde \rho}(p',q')
\eqno(6.9)
$$
is bounded from below by $I=1$, with equality if and
only if $ \tilde \rho$ is a coherent state. Now the point is that (6.8)
and (6.9) have a very similar form: they are both Wigner functions of
an arbitrary density matrix with a Gaussian smearing, but the Gaussian
factors are not the same.  Our aim, therefore, is to perform a
series of transformations to bring (6.8) into the form (6.9), and then
apply the Lieb-Wehrl result (2.16). As we shall see, the information of
$\mu$ is not preserved under these transformations, and thus we obtain
a non-trivial lower bound, different from (2.16), and depending on the
quantity $A_z^t$.  The difficulty outlined above is avoided because
the evolution under $J$ is contained entirely in $A_t^z$, and the
minimization is now over all pure $\rho_0$, a well-defined class to which
the Lieb-Wehrl result may be applied.

Turn now to the details. Consider first Eq.(6.7).
The final density matrix is
$$
\la x | z \ra \la z | y \ra = {1 \over (2 \pi \s_q^2)^{\half} }
\ \exp \left( - { \xi^2 \over 8 \s_q^2 }
- {(X-2 \q)^2 \over 8 \s_q^2 } + \ih \p \xi \right)
\eqno(6.10)
$$
where as in Section IV, $X= x+y$, $ \xi = x-y $.
Under evolution backwards in time by the non-unitary propagator $J$
it yields,
$$
A_t^z(y_0,x_0) = {N \over L }
\left( { 2 \alpha_0 \over \pi } \right)^{\half}
\ \exp \left( - \a_0 (X_0 - 2 \q_0)^2 - \b_0 \xi_0^2
+ \ih \xi_0 \left[ \G_0 (X_0 - 2 \q_0) + \p_0 \right] \right)
\eqno(6.11)
$$
where
$$
\eqalignno{
\a_0 &= { L^2 \over 32 \s_q^2 \Delta }
&(6.12)\cr
\b_0 &= C + 2 \s_q^2 N^2 - { ( B - 4  \s_q^2 N \tilde K)^2 \over
32 \s_q^2 \Delta }
&(6.13) \cr
\G_0 &= \hat K + {L \over 4 \Delta} \left( {B \over 4 \s_q^2}
- N \tilde K \right)
&(6.14) \cr }
$$
Here,
$$
\Delta = { 1 \over 8 \s_q^2 } \left( A + {1 \over 8 \s_q^2} \right)
+ {1 \over 4} {\tilde K}^2
\eqno(6.15)
$$
(These coefficients may be obtained by a straightforward modification of
the calculations described in Ref.[\cite{dowker}]).
Also, $\p_0, \q_0$ are the classical evolution of $\p, \q$, evolved
backwards in time under the dissipative equation of motion (4.10). They are
given explicitly by
$$
\eqalignno{
\p_0 &= - {2 \over L} ( N L - \hat K \tilde K ) \q +
{ \hat K \over L} \p
&(6.16) \cr
\q_0 &= { \tilde K \over L} \q + { 1 \over 2 L } \p
&(6.17) \cr }
$$
where the various quantities appearing are defined in Section IV.
This transformation from $\p,\q$ to $\p_0, \q_0$ is non-canonical,
because the evolution is dissipative:
$$
{ \partial (\q_0,\p_0 ) \over \partial (\q, \p) } = {N \over L}
= e^{2\gamma t}
\eqno(6.18)
$$
Performing the Wigner transform, one thus obtains the explicit form
of (6.8):
$$
\eqalignno{
\mu_t(\p,\q) = 2 {N \over L }
\left( {\a_0 \over \b_0 } \right)^{\half}
\int dp dq \exp & \left( - {1 \over 4 \hbar^2 \b_0} ( p - \p_0
- 2  \G_0 ( q - \q_0))^2 - 4 \a_0 (q- \q_0)^2 \right) \cr
\times & \ W_{\rho_0}(p,q)
&(6.19) \cr}
$$
We would like to bring this expression into the form (6.9). Introduce
$$
\lambda = \left( { \b_0 \over \a_0 } \right)^{\half}, \quad \quad
\mu = \sqrt{8} \s_q (\b_0 \a_0)^{1/4}
\eqno(6.20)
$$
Now perform the following canonical transformation on the integration
variables, together with the same change of variables on $\p_0, \q_0$:
$$
\eqalignno{
q' &= \mu  q, \quad \quad
p' = {1 \over \mu} (p - 2 \G_0 q)
&(6.21) \cr
\tilde q &= \mu \q_0, \quad \quad
\tilde p  = {1 \over \mu} (\p_0 - 2 \G_0 \q_0)
&(6.22) \cr }
$$
Eq.(6.19) thus becomes,
$$
\mu_t (\p, \q) =  2 { N \over L }
\left( {1 \over \lambda } \right)^{\half}  \int dp' dq'
\ \exp \left( - { (p' - \tilde p)^2 \over 2 \lambda \sigma_p^2 }
-{ (q' - \tilde q)^2 \over 2\lambda \sigma_q^2 } \right)
\ W_{\tilde \rho} (p',q')
\eqno(6.23)
$$
where we have introduced
$$
W_{\tilde \rho} (p',q') =
W_{\rho_0}(\mu p' + 2 {\G_0 \over \mu} q',{q'\over \mu})
\eqno(6.24)
$$
There arises the question of whether $W_{\tilde \rho}(p',q')$ defined by (6.24)
is still a
Wigner function, {\it i.e.}, of whether there exists a density matrix
$\tilde \rho$ whose Wigner transform is (6.24).  The answer is in the
affirmative:  linear canonical transformations on the arguments of the
Wigner function are readily shown to correspond to unitary
transformations of $\rho$.

The dependence on $\p,\q$ in the
right-hand side of (6.23) resides entirely in $\tilde p, \tilde q$, via
the transformations (6.16), (6.17) and (6.22). It is convenient to
write (6.23) as
$$
\mu_t(\p, \q) = {N \over L} \tilde \mu_t (\tilde p, \tilde q)
\eqno(6.25)
$$
The factor ${N \over L}$ is nothing
more than the Jacobean of  the transformation from $\tilde p,
\tilde q $ to $\p$, $\q$.
The transformation (6.21), (6.22) is canonical so the
only contribution to the Jacobean comes from (6.16), (6.17), whose
Jacobean is (6.18).
The information of $\mu_t$, $I_t$, is then simply related to that of $\tilde
\mu_t$, $\tilde I_t$.
It is
$$
I_t = \tilde I_t - \ln \left( {N \over L} \right)
\eqno(6.26)
$$

The distribution $\tilde \mu_t $
is almost of the desired form (6.9), but fails to be
because of the presence of the factor of $\lambda$. Positivity of the
density matrix (6.11), implies that $\b_0 \ge \a_0$, and in fact
equality holds only at $t=0$, and thus one has $\lambda > 1 $.
One might have thought that the next step is to simply scale $p'$ and
$q'$ by $\lambda^{\half}$, thus taking $\lambda$ into the Wigner
function. However, this scaling would lead to a phase space
distribution function which is {\it not} a Wigner function, {\it i.e.},
it is not the Wigner transform of a density matrix. This is easy to
see:  under such a scaling,
the degree to which the Wigner function may be peaked about a
region of phase space becomes enhanced by a factor of $\lambda > 1$,
and thus it is possible to violate the uncertainty principle. Wigner
functions scaled in this way cannot therefore correspond to density
matrices.

Instead, the next step is carried out using the following simple fact
about convolution integrals: when two Gaussians with variances
$\s_1$ and $\s_2$ are convoluted, the variance of their convolution,
$\s_3$, satisfies $\s_3^2 = \s_1^2 + \s_2^2 $. Let us therefore
express the Gaussian smearing function in $\tilde \mu_t$ as the convolution
of two Gaussians:
$$
\eqalignno{
{1 \over \pi \hbar \lambda}
\ \exp \left( - { (p'-\tilde p)^2 \over 2 \lambda \sigma_p^2 }
-   { (q'- \tilde q )^2 \over 2 \lambda \sigma_q^2 } \right)
& =   \int   dp dq  \ {1 \over \pi \hbar}
\ \exp  \left( - { (p -  p')^2 \over 2  \sigma_p^2 }
-{ (q - q')^2 \over 2 \sigma_q^2 } \right)
\cr
\times \ { 1 \over \pi \hbar (\lambda -1) }
\  \exp & \left( - { (p - \tilde p)^2 \over 2 (\lambda -1)\sigma_p^2 }
-{ (q - \tilde q)^2 \over 2(\lambda  -1)\sigma_q^2 } \right)
&(6.27) \cr }
$$
We may therefore write $\tilde \mu_t $  as
$$
\tilde \mu_t (\tilde p, \tilde q) = { 1 \over \pi \hbar (\lambda -1)}
\ \int dp dq
\ \exp \left( - { (p - \tilde p)^2 \over 2 (\lambda -1)\sigma_p^2 }
-{ (q - \tilde q)^2 \over 2(\lambda -1)\sigma_q^2 } \right)
\ \hat \mu_t (p,q)
\eqno(6.28)
$$
where $\hat \mu_t $ is precisely of the form (6.9), with Wigner function
$W_{\tilde \rho} (p',q')$, given above by (6.24).

The result (6.28) is as close as we can get to casting $\mu(\p,\q)$
in the form (6.9). However, the form (6.28) may be exploited:
it is the convolution of a Gaussian
with the function $\hat \mu_t$. We may therefore appeal to a theorem
of Lieb on the information
of convolutions [\cite{lieb}]. Let $f$ and $g$ be functions defined in
$L^s(R^n)$,
where $s>1$, and let $f*g$ denote their convolution.
Then the information of $f*g$, $I(f*g)$, satisfies the inequality,
$$
\exp \left( {2 \over n} I(f*g) \right)
\ \ge \
\exp \left( {2 \over n} I(f) \right) + \exp \left( {2 \over n} I(g) \right)
\eqno(6.29)
$$
Equality holds if $f$ and $g$ are both Gaussians differing only in
the location of their centres and in an overall scale of their covariance
matrices.

In our case, the Gaussian function in (6.28) has information
$$
I = \ln \left( { e \over 2 } ( \lambda -1 ) \right)
\eqno(6.30)
$$
Since $\hat \mu_t$ is of the form (6.9), it satisfies the Lieb-Wehrl
inequality (2.16), with equality if and only if the Wigner function
(6.24) is the Wigner function of a coherent state,
$$
W_{\rho_0}(\mu p' + 2 {\G_0 \over \mu} q',{q'\over \mu} )
= \ {1 \over \pi \hbar}
\ \exp  \left( - { (p' -  p^{\prime}_1)^2 \over 2  \sigma_p^2 }
-{ (q' - q^{\prime}_1)^2 \over 2 \sigma_q^2 } \right)
\eqno(6.31)
$$
Applying (6.29),
we therefore have the lower bound on the information of $\tilde \mu_t$,
$$
\tilde I_t \ \ge \ 1 + \ln \left( \half ( \lambda + 1 ) \right)
\eqno(6.32)
$$
Finally, inserting this in (6.26),we obtain
the desired lower bound on $I_t$:
$$
I_t \ \ge \  1 + \ln \left( {L \over 2 N} \left[
\left( {\b_0 \over \a_0} \right)^{\half} + 1 \right]\right)
\eqno(6.33)
$$
This is our main result. The right-hand side is the value of $I_t^{min}$
at time $t$.

Now consider the conditions for equality in (6.33), to determine the
initial state which meets the envelope at time $t$. The information
of $\hat \mu_t$ achieves its lower bound when (6.31) is satisfied.
$\hat \mu_t$ is then a Gaussian, differing only from the smearing
Gaussian in (6.28) by an overall scaling of their covariance matrices.
The conditions for equality in (6.29) are therefore also satisfied.
This means that the inequality (6.33) achieves equality when the
initial state is given by (6.31). Inverting the Wigner transform,
we find that the initial state is the pure state,
$$
\Psi_t(x) =  \left( { 4 ( \a_0 \b_0)^{1/2} \over \pi } \right)^{1/4}
\ \exp \left(  - [2 (\a_0 \b_0)^{\half} + \ih \G_0 ] (x-\q_1)^2 +
\ih \p_1 x \right)
\eqno(6.34)
$$
This is a squeezed coherent state.

\def\p{{\bar p}}
\def\q{{\bar q}}

\head {\bf VII. Discussion and Summary}

We first discuss the properties of the lower bound (6.33).

Consider Eq.(6.19). We have been seeking the Wigner function $W_{\rho_0}$
that minimizes the information of (6.19). Loosely speaking, this means
finding the Wigner function which has the best overlap with the
exponential in (6.19), and hence gives the most peaked probability
distribution $\mu_t(\p,\q)$.
We found that the initial state doing the
job is (6.34), whose Wigner transform is, from (6.31),
$$
W_{\Psi}(p,q) = { 1 \over \pi \hbar} \ \exp \left( -
{( p - 2 \G_0 q - \mu p^{\prime}_1)^2
\over 2 \mu^2 \s_q^2 } - { \mu^2 \over 2 \s_q^2} \left(q - {q^{\prime}_1
\over \mu} \right)^2 \right)
\eqno(7.1)
$$

Now consider the exponential function in (6.19). It is the Wigner function
of the final coherent state evolved backwards by $J$. The contours
of the Wigner function start out as circles. Each contour suffers three
effects under this non-unitary evolution:
it is distorted into an ellipse, its
axes are rotated, and its area increases. The distortion factor
is given by $\mu$ in (6.20), the amount of rotation is given by $\G_0$,
and the area increase is given by $\lambda$. (There is in addition a
translation of the contours, but this preserves
the information.)

Now the point is that the Wigner function (7.1) giving the least
overlap in (6.19) is the Gaussian pure state which matches two out of
three of these effects: it has the same distortion and rotation
factors.  It does not have the same expansion factor $\lambda$ -- it
cannot because we know that the minimizing state must be pure, and pure
Gaussian states must have $\lambda =1 $. The minimizing state is
therefore the state whose Wigner function is close as posible to the
exponential in (6.19) subject to the constraint that it be pure.

Turn now to the explicit form of the lower bound. Using the result of
Refs.[\cite{dowker,huzhang,hupaz,caldeira}], it may be shown that in
the Fokker-Planck limit, and
for short times, one has
$$
\left( \b_0 \over \a_0 \right)^{\half}
\approx 1 + 2 \gamma t + { 8 \s_q^2 M \gamma k T \over \hbar^2} t
+ O(t^2)
\eqno(7.2)
$$
Setting $\s_q$ to the value (5.6), one thus has
$$
I_t^{min} = 1 +
\ln \left[ 1 + \left( { 2 k T \over \hbar \omega_R} - 1 \right)
\gamma t + O(t^2) \right]
\eqno(7.3)
$$
The Fokker-Planck limit involves $kT >> \hbar \omega_R $, so $I_t^{min}$
increases with time.  Eq.(7.3) indicates that the thermal contributions
to the uncertainty principle start to become appreciable after
a time
$$
t \sim { \hbar^2 \over \s_q^2 M \gamma k T } \sim { \hbar \omega_R
\over \gamma k T }
\eqno(7.4)
$$
The important thing to note is that this is the decoherence timescale
defined in Eq.(5.7) -- the timescale on which interference is
destroyed by the interaction with the environment.

Our results should be compared with the work of
Hu and Zhang [\cite{huzhang}]. They calculated the usual uncertainty function,
(3.10), for the density matrix obtained by evolving an initial
coherent state for time $t$ in the presence of an environment.
They found that for short times, and high temperatures,
$$
U = { \hbar^2 \over 4 }
\left[ 1 + \left( { 2 k T \over \hbar \omega_R} - 1 \right)
\gamma t + O(t^2) \right]
\eqno(7.5)
$$
It was these authors who first noted, on the basis of this
calculation, the significance of the decoherence timescale for
the comparative sizes of thermal and quantum fluctuations. We
thus find close agreement with their work.

This result has a consequence for the decoherence programme.
A reasonable question to ask in decoherence models is whether there
is a regime in which the interaction with the environment is
sufficient to induce decoherence, yet induces a noise level less
than that due to intrinsic quantum fluctuations. Our results,
and those of Hu and Zhang [\cite{huzhang}], show that this is not
the case: in the Fokker-Planck regime, decoherence and thermal
fluctuations become important on the same timecale. This means,
loosely speaking, that the uncertainty principle plays little role
in these models.

It is of interest to explore the form of the lower bound in other
regimes. Consider for example, the low temperature regime. In
the Fokker-Planck (high temperature) regime discussed above, the
diffusion is controlled by the diffusion constant $D = 2 M \gamma
kT$. However, as argued by Caldeira and Leggett [\cite{caldeira}],
in the low temperature regime the appropriate diffusion constant is
$D = M \gamma \hbar \omega_R $. An order of magnitude estimate on
the size of $I_t^{min}$ is therefore obtained by substitution of
diffusion constants. One thus discovers that in the low temperature
regime, the environmentally-induced fluctuations (we can no longer
call them thermal) grow on a timescale $\gamma^{-1}$, the relaxation
timescale. This shows that $I_t^{min}$ is not just a measure of
quantum fluctuations of the distinguished system plus thermal
fluctuations of the environment: it also includes the quantum
fluctuations of the environment (although these are of course
negligible in the high temperature regime).

Another question to ask is whether it is possible to express our new
uncertainty principle (6.33) in terms of the usual uncertainty function
$U$. Recall that the Lieb-Wehrl inequality (2.16) may be shown to imply
the standard uncertainty principle, (1.1), via the steps
(2.17)--(2.20). Can a similar derivation be carried out in the case of
(6.33)? Steps
analagous to (2.17)--(2.20) can be carried out, and one obtains,
$$
\Delta_{\mu} p \Delta_{\mu} q \ \ge \ { L \over 2 N}
\left[ \left( { \b_0 \over \a_0} \right)^{\half} + 1 \right] \hbar
\eqno(7.6)
$$
However, as before this is not the proper form of the uncertainty principle,
because the variances on the left-hand side also include the variances
of the coherent state, Eqs.(2.18), (2.19).  The final step of
minimizing over $\s_q$ is rather tricky to carry out because the
right-hand side of (7.6) depends on $\s_q$ in a non-trivial way,
and one ends up with a fifth order polynomial in $\s_q^2$.
Also, the alternative method suggested by Grabowski
[\cite{grabowski}] cannot
obviously be generalized so as to apply to this case.
Therefore, we do not give an explicit form of our
uncertainty relation in terms of the variances of $\rho$.
The possibility of deriving such a relation directly (rather
than from (7.6)) will be considered elsewhere [\cite{jjh3}].

We should also compare with the work of Paz, Habib and Zurek
[\cite{zurekpazhabib,paz}], who looked for the set of initial states
which generated the least amount of physical entropy, $S[\rho]$,
after evolution in the presence of an environment for time $t$.
The motivation for doing this is that these states are
in a sense the ones most stable under evolution in the presence
of an environment. This is clearly closely related to
our work, since we essentially looked for the set of initial states
with the smallest value of $I$ after time $t$. Indeed, Paz {\it et al.}
claimed that the minimizing states are coherent states, whereas
for us the minimizing states are squeezed coherent states. It turns
out that the quantity controlling the squeezing, $\Gamma_0$ in
(6.34), can go to zero quite quickly (on the timescale
$\omega^{-1}$). In this case we thus see that the results are
in agreement.

In summary, we have discussed the properties of an
information-theoretic measure of uncertainty (1.3) for a class of
non-equilibrium quantum systems. Our measure is closely related to
von Neumann entropy in the thermal regime, but unlike entropy, it
supplies a non-trivial measure of uncertainty in the quantum regime.
It is also easier to work with calculationally than entropy. Our
main result is the demonstration that, for linear systems,  our
measure has a non-trivial lower bound, the generalization of the
uncertainty principle to include thermal (or more generally,
environmentally-induced) fluctuations for a class of non-equilibrium
systems. We have examined the form of the lower bound in some
regimes of interest. A more detailed examination is best carried out
numerically, but this is beyond the scope of the present work.

\head {\bf Acknowledgements}

We would like to thank many of our colleagues for useful
conversations, including Carlton Caves, Murray Gell-Mann,
Salman Habib, Jim Hartle, Chris Isham,  Raymond
Laflamme, Seth Lloyd, Juan Pablo Paz and Wojtek Zurek. One of us
(J.J.H.) would like to thank Wojtek Zurek for his invitation to
visit the Los Alamos National Lab, where part of this work was
carried out, and Bei-Lok Hu, for hosting a visit to the University
of Maryland. We are particularly grateful to Bei-Lok Hu for useful
conversations and for encouraging us to pursue these ideas. His
paper with Zhang [\cite{huzhang}] was the inspiration for this piece
of work.

A.A. was supported by the Science and Engineering Research Council.
J.J.H. was supported by a University Research Fellowship from the
Royal Society.

\references

\def\pr{{\sl Phys. Rev.\ }}

\def\prep{{\sl Phys. Rep.\ }}

\def\rmp{{\sl Rev. Mod. Phys.\ }}
\def\cmp{{\sl Comm. Math. Phys.\ }}

\def\annp{{\sl Ann. Phys. (N.Y.)\ }}

\def\jsp{{\sl J. Stat. Phys.\ }}

\refis{abe} S.Abe and N.Suzuki, \pr {\bf A41}, 4608 (1990).

\refis{anderson} A.Anderson, \pr {\bf D42}, 585 (1990).

\refis{caldeira} A.O.Caldeira and A.J.Leggett, {\sl Physica} {\bf
121A}, 587 (1983).

\refis{dekker} H.Dekker, \pr {\bf A16}, 2116 (1977).

\refis{dowker} H.F.Dowker and J.J.Halliwell, \pr {\bf D46}, 1580
(1992).

\refis{gellmann}
M.Gell-Mann and J.B.Hartle, in {\it Complexity, Entropy and the
Physics of Information. SFI Studies in the Sciences of Complexity,
Vol. VIII}, edited by W.Zurek (Addison Wesley, Reading, MA, 1990);
M.Gell-Mann and J.B.Hartle, ``Classical Equations for Quantum
Systems'', UCSB preprint (to be published in \pr D, 1993);
J.B.Hartle, in {\it Quantum Cosmology and Baby Universes:
Proceedings of the 1989 Jerusalem Winter School on Theoretical
Physics}, edited by S.Coleman, J.B.Hartle, T.Piran and S.Weinberg
(World Scientific, Singapore, 1991); and in ``Spacetime Quantum
Mechanics and the Quantum Mechanics of Spacetime'', preprint
UCSBTH92-91 (to appear in proceedings of the 1992 Les Houches Summer
School, {\it Gravitation et Quantifications}).

\refis{griffiths} R.Griffiths, \jsp {\bf 36}, 219 (1984).

\refis{grabowski} M. Grabowski, {\sl Rep.Math.Phys.} {\bf 20}, 153 (1984).

\refis{grabert} H.Grabert, P.Schramm, G-L. Ingold, \prep {\bf 168},
115 (1988).

\refis{hupaz} B.L.Hu, J.P.Paz and Y.Zhang, \pr {\bf D45}, 2843
(1992); ``Quantum Brownian Motion in a General Environment. II:
Non-Linear Coupling and Perturbative Approach'', Los Alamos Preprint
lA-UR-92-1367 (1992).

\refis{husimi} K.Husimi, {\sl Proc.Phys.Math.Soc. Japan} {\bf 22},
264 (1940).

\refis{huzhang} B.L.Hu and Y.Zhang, ``Uncertainty Relation at Finite
Temperature'', University of Maryland Preprint (1992).

\refis{info} The foundations of information theory (then known as
communcation theory) were set out by C.Shannon in,  {\it The
Mathematical Theory of Communication}, C.E.Shannon and  W.W.Weaver
(University of Illinois Press, Urbana, IL, 1949). For a more recent
account, in which the basic results used here are described, see
for example, {\it Elements of Information Theory}, T.M.Cover and
J.A.Thomas (Wiley, New York, 1991). One of the first to apply
information theory to quantum mechanics was H.Everett. See his
article in, {\it The Many-Worlds Interpretation of Quantum
Mechanics}, edited by B.S.DeWitt and N.Graham (Princeton University
Press, Princeton, NJ, 1973).

\refis{jjh} J.J.Halliwell, \pr {\bf D46}, 1610 (1992).

\refis{jjh2} J.J.Halliwell, ``Quantum-Mechanical Histories and
the Uncertainty Principle.
I. \ Information-Theoretic Inequalities'',
Imperial College Preprint (1993).

\refis{jjh3} J.J.Halliwell, in prepration.

\refis{klauder} J.R.Klauder and E.C.G.Sudarshan, {\it Fundamentals
of Quantum Optics} (Benjamin, New York, NY, 1968); J.R.Klauder and
B.S.Skagerstam, {\it Coherent States} ( World Scientific, Singapore,
1985).

\refis{lieb} E.H.Lieb, \cmp {\bf 62}, 35 (1978).

\refis{omnes} R.Omn\`es, \rmp {\bf 64}, 339 (1992), and references
therein.

\refis{paz} J.P.Paz, S.Habib and W.Zurek, \pr {\bf D47}, 488 (1993).

\refis{unruh} W.G.Unruh and W.H.Zurek, \pr {\bf D40}, 1071 (1989).

\refis{vernon} R.P.Feynman and F.L.Vernon, \annp {\bf 24}, 118 (1963).

\refis{wehrl} A.Wehrl, {\sl Rep.Math.Phys.} {\bf 16}, 353 (1979).

\refis{wigner} N.Balazs and B.K.Jennings, \prep {\bf 104}, 347 (1984),
M.Hillery, R.F.O'Connell, M.O.Scully and E.P.Wigner, \prep {\bf
106}, 121 (1984).

\refis{zeh} E.Joos and H.D.Zeh, {\sl Z.Phys.} {\bf B59}, 223 (1985).

\refis{zurek} W.Zurek, in {\it Physical Origins of Time Asymmetry},
edited by J.Halliwell, J.Perez-Mercader and W.Zurek (Cambridge
University Press, Cambridge, to appear, 1993).

\refis{zurekpazhabib} W.Zurek, S.Habib and J.P.Paz, ``Coherent
States via Decoherence'', Los Alamos Preprint LA-UR-92-1642 (1992).

\refis {zurek2} W.H.Zurek, in {\it Frontiers of Non-Equilibrium
Statistical Mechanics}, edited by G.T.Moore and M.O.Scully (Plenum,
New York, 1986).

\endreferences

\end